\documentclass[
 reprint,
 superscriptaddress,
 amsmath,
 amssymb,
 aps,
 prl,
]{revtex4-2}

\usepackage{graphicx}
\usepackage{dcolumn}
\usepackage{bm}

\usepackage{subcaption}
\usepackage[justification=raggedright,singlelinecheck=false]{caption}
\usepackage[thinc]{esdiff} 
\usepackage{xcolor}
\usepackage[colorlinks=true, allcolors=blue]{hyperref}

\def\ie{{\textit{i.e.} }}

\def\to{\rightarrow}

\usepackage{afterpage}

\begin{document}

\preprint{APS/123-QED}

\title{Activation and Avalanche Length Scales in the Finite-Temperature Creep of an Elastic Interface}

\author{Giovanni Russo}
\affiliation{%
Université Paris-Saclay, CNRS, LPTMS, 91405, Orsay, France.
}%
\affiliation{%
PMMH, CNRS, ESPCI Paris, Université PSL, Sorbonne Université, Université
Paris Cité, France.
}%
\email{Contact author: giovanni.russo@universite-paris-saclay.fr}%

\author{Ezequiel E. Ferrero}
\affiliation{
 Instituto de Nanociencia y Nanotecnología, CNEA–CONICET,\\
 Centro Atómico Bariloche, (R8402AGP) San Carlos de Bariloche, Río Negro, Argentina.
}%

\author{Alejandro B. Kolton}
\affiliation{%
 Instituto Balseiro, Centro Atómico Bariloche, CNEA–CONICET–UNCUYO,
R8402AGP San Carlos de Bariloche, Río Negro, Argentina. 
}%

\author{Alberto Rosso}%
\affiliation{%
Université Paris-Saclay, CNRS, LPTMS, 91405, Orsay, France.
}%

\author{Damien Vandembroucq}%
\affiliation{%
PMMH, CNRS, ESPCI Paris, Université PSL, Sorbonne Université, Université
Paris Cité, France.
}%


\begin{abstract}
We investigate the creep dynamics of a driven elastic line at finite temperature, well below the depinning threshold. We show that creep is governed by two distinct length scales. The first, $\ell_{\mathrm{opt}}$, corresponds to the optimal activated rearrangements that control the dynamics' bottleneck and remains essentially temperature-independent. The second, $\ell_{\mathrm{av}}$, characterizes the spatial extent of thermal avalanches and grows as temperature decreases. By combining structural and dynamical observables, we show that $\ell_{\mathrm{av}}$ governs both the crossover in the structure factor and the growth of the four-point dynamical susceptibility, while the relaxation time remains controlled by activation over large barriers associated with $\ell_{\mathrm{opt}}$. We find that the avalanche scale follows $\ell_{\mathrm{av}}(T)\sim T^{-\nu_{\mathrm{dep}}}$, thereby selecting a unique scenario among competing theoretical predictions. These results establish a unified picture of finite-temperature creep in which activation controls temporal scales while depinning criticality governs spatial correlations.
\end{abstract}

\maketitle


\begin{figure*}[htbp]
\centering
\begin{subfigure}{0.49\textwidth}
    \centering
    \includegraphics[width=\textwidth]{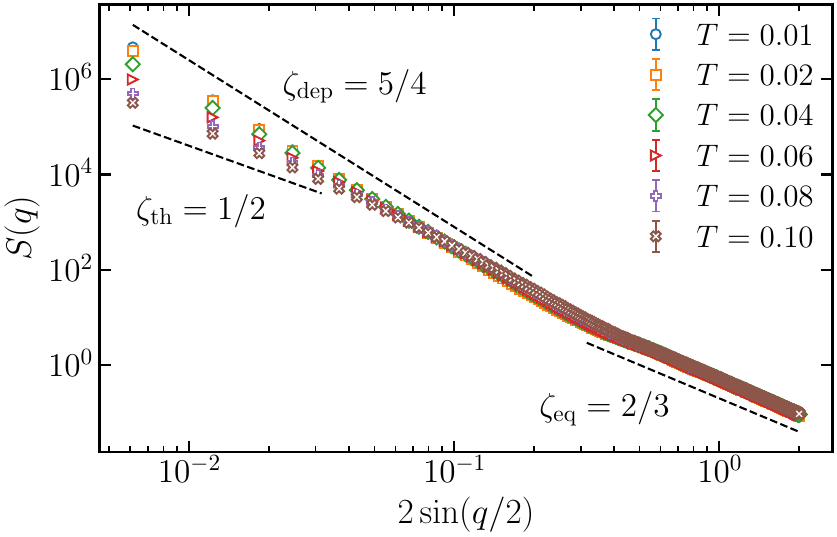}
\end{subfigure}
\hfill
\begin{subfigure}{0.49\textwidth}
    \centering
    \includegraphics[width=\textwidth]{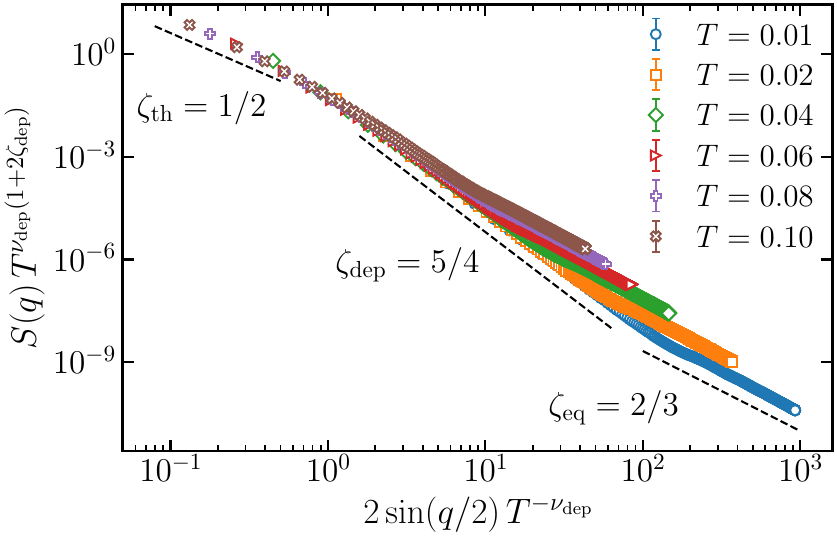}
\end{subfigure}
\caption{
Structure factor $S(q)$ for a line of size $L=1024$ at fixed drive $f/f_c = 0.1$. 
For each temperature, the structure factor is averaged over $10^4$ configurations.
Left: raw data for different temperatures. The crossover between equilibrium and depinning regimes, set by the optimal rearrangement scale $\ell_{\mathrm{opt}}(f)$, remains essentially unchanged with temperature. At larger length scales, a second crossover emerges as temperature increases, where the small-$q$ sector progressively approaches the thermal roughness exponent $\zeta_{\mathrm{th}}=1/2$. 
Right: data collapse obtained by rescaling wavevectors with $T^{-\nu_{\mathrm{dep}}}$, demonstrating that this temperature-dependent crossover defines a length scale $\ell_{\mathrm{av}}(T)\sim T^{-\nu_{\mathrm{dep}}}$.
}
\label{fig:fig1}
\end{figure*}

Most of the time, motion is almost imperceptible: a granular column settles under load~\cite{Amon-PRL12}; a gel deforms under shear~\cite{Divoux-PRL14, Ramos-PRL20}; a concrete wall deforms under compression~\cite{Ulm-PNAS09}; a sheet of paper is damaged under tension~\cite{Rosti2010}; the blocks of a geological fault slip past each other~\cite{Wesson-JGR88}; crease by crease a crumpled thin sheet unfolds~\cite{Lahini-PRL23}; day after day, a fissure propagates across glass~\cite{Wiederhorn-JACS67, Guin-JNCS03}; magnetic domain walls drift through a disordered medium~\cite{Lemerle-PRL98}. Yet this apparent smoothness conceals intermittent dynamics in both the transient and steady-state regimes, with sudden events separated by long waiting times. Understanding the origin of this intermittency and the associated length scales remains a central challenge in thermally activated dynamics.

Driven disordered dynamics arises from the interplay of elasticity, quenched disorder, and thermal fluctuations. Two extensively studied classes of systems are elastic interfaces in random media~\cite{Fisher1998, Kardar1998, Ferrero2021} and amorphous solids with long-range Eshelby-mediated elastoplastic interactions~\cite{Nicolas2018, Bouttes2013creep, Baret2002, Ozawa2023}. In the athermal limit, both systems exhibit a threshold separating arrested and flowing regimes: the depinning threshold $f_c$ for interfaces and the yielding threshold $\Sigma_Y$ for amorphous materials, whose finite-temperature flow has been extensively investigated~\cite{Haxton2007, Chattoraj2010, Karmakar2010FiniteT, Popovic2021b, Ferrero2021Yielding, Korchinski2022}. When driven below these thresholds, motion occurs via thermally activated processes, resulting in extremely slow dynamics known as creep.
In the conventional picture of creep, rare activated events allow the system to overcome large energy barriers separating metastable configurations. This framework successfully accounts for the average velocity through Arrhenius-like laws and identifies a characteristic activation scale~\cite{Ioffe1987, Vinokur1996, Kolton2005}, but does not capture the spatially heterogeneous character of the dynamics. Such dynamical heterogeneities, widely studied in glassy systems~\cite{Berthier2011}, have also been observed in the creep dynamics of disordered elastic systems~\cite{Chauve2000, Ferrero2017, Durin2024, Tahaei2023}, pointing to collective relaxation processes beyond the standard activation picture.
More recently, attention has focused on correlated collective events called thermal avalanches. Numerical studies in the limit of vanishing temperature ($T \to 0^+$) have shown that successive activated events organize into correlated cascades extending far beyond the individual excitation scale~\cite{Kolton2009, Vandembroucq2004, Ferrero2017}, while experiments have directly observed avalanche-like creep with strong spatiotemporal correlations~\cite{Grassi2018, Durin2024, Lahini2023}.

Despite these advances, how finite temperature controls the avalanche extent $\ell_{\mathrm{av}}$ remains poorly understood. A central question is whether finite-temperature creep is governed by a single scale or by distinct mechanisms controlling temporal and spatial properties.

We study finite-temperature creep in a one-dimensional directed polymer, a minimal model of a driven elastic interface in a disordered medium. It represents several experimentally relevant one-dimensional interfaces~\cite{Wiederhorn-JACS67, Guin-JNCS03, Lemerle-PRL98} and retains the essential ingredients of creep: elasticity, quenched disorder, metastability, thermal activation, and avalanche-like relaxation. At the same time, it provides a controlled setting with accurately known equilibrium and depinning exponents. A Dijkstra-based search~\cite{Dijkstra1959} gives direct access to low-energy excitations, enabling simulations deep in the creep regime.

Our results show that finite-temperature creep is governed by two distinct length scales. The first, $\ell_{\mathrm{opt}}(f)$, is the optimal activation
scale. It sets the equilibrium-to-depinning crossover in the structure factor, $S(q)$, and the size of the rearrangement associated with the dominant activation barrier. We show that this scale remains temperature-independent and that at small length scales the interface retains its equilibrium roughness. The second, $\ell_{\mathrm{av}}(f,T)$, is the avalanche scale. It sets the depinning-to-thermal crossover in $S(q)$ and the spatial extent of the cascade of events that follows the initial activation. By analyzing the structure factor~\cite{Ferrero2019} and the four-point dynamical susceptibility~\cite{Berthier2011,Tahaei2023}, we show that, for fixed $f$,
\begin{equation}
\ell_{\mathrm{av}}(f,T) \sim T^{-\nu_{\mathrm{dep}}},
\end{equation}
where $\nu_{\mathrm{dep}}$ is the depinning correlation-length exponent. This result discriminates among competing theoretical predictions for the avalanche cutoff~\cite{Chauve2000, Chauve2001, Tahaei2023,deGeus2024}, favoring the scenario proposed in Ref.~\cite{deGeus2024}. Moreover, it extends its validity to the deep collective creep regime, characterized by droplet scaling and ultraslow dynamics driven by diverging energy barriers.

Our results thus provide a unified picture of creep dynamics in which activation controls temporal scales, while depinning criticality governs spatial correlations. 
Despite the different interaction kernels and universality classes of depinning and yielding, this separation may provide a useful framework for thermally activated dynamics in amorphous solids, where the sign-changing Eshelby kernel is associated with a pseudogap in local stabilities and the existence of a counterpart of $\ell_{\mathrm{opt}}$ remains an open question~\cite{MullerWyart2015, Zippelius2023}.

\begin{figure*}[htbp]
\centering

\begin{minipage}[t]{0.49\textwidth}
    \centering
    \includegraphics[width=\textwidth]{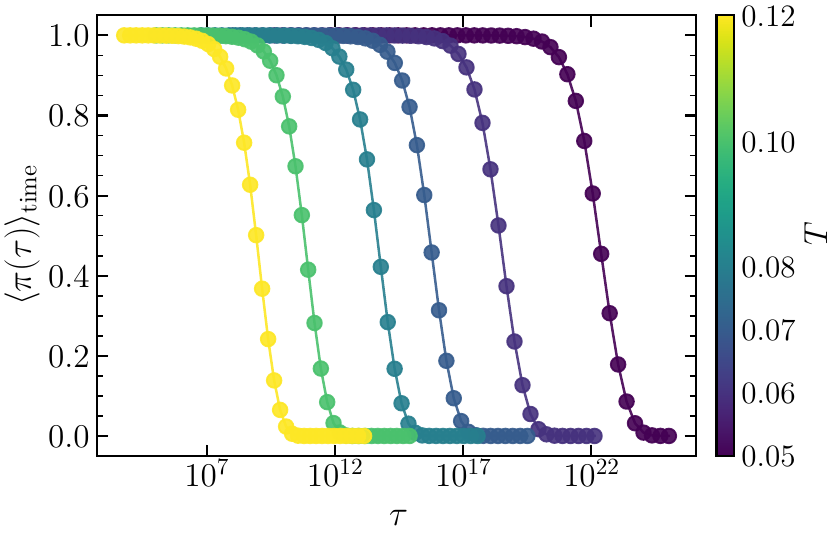}
\end{minipage}
\hfill
\begin{minipage}[t]{0.49\textwidth}
    \centering
    \includegraphics[width=\textwidth]{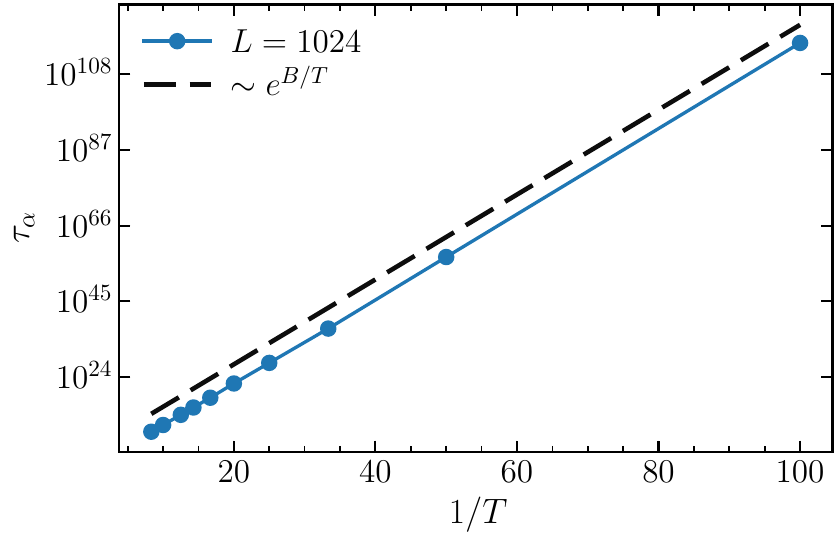}
\end{minipage}

\caption{
Relaxation dynamics at finite temperature. 
Left: Persistence $\langle \pi(\tau)\rangle_\mathrm{time}$ at fixed drive $f/f_c = 0.1$ and system size $L=512$ for different temperatures. As temperature decreases, relaxation slows down and becomes increasingly heterogeneous, as evidenced by the slower decay of persistence. 
Right: relaxation time $\tau_\alpha$, defined by $\langle \pi(\tau_\alpha)\rangle = 0.5$, as a function of inverse temperature for $L=1024$. The exponential growth of $\tau_\alpha$ with $1/T$ demonstrates Arrhenius behavior, indicating that the dynamics is controlled by thermally activated processes over large, temperature-independent energy barriers.
}
\label{fig:fig2}
\end{figure*}

\emph{The model.} Under a weak drive, an elastic interface in a disordered medium evolves through rare thermally activated excitations out of metastable states, followed by deterministic relaxation toward new, lower-energy metastable configurations~\cite{Kolton2005}. In a lattice representation, these excitations can be explicitly identified to construct an effective dynamics~\cite{Kolton2006, Kolton2009}. This construction was developed for $T\to0^+$, where the optimal activated event dominates~\cite{Ferrero2017}. Here, we extend it to finite temperature.

We consider a directed polymer of $L$ monomers with integer transverse displacements $u(i)$, $i=0,\dots, L-1$, and periodic boundary conditions. Its energy reads
\begin{equation}
E=\sum_{i=0}^{L-1} \left\{\frac{1}{2}\left[u(i+1)-u(i)\right]^2 - f\,u(i) + V(i,u(i))\right\},
\end{equation}
where $f$ is the drive and $V(i,u)$ is uncorrelated Gaussian disorder. 
For computational convenience, we impose the metric constraint $|u(i+1)-u(i)|\le K$ and set $K=1$. This choice is discussed in the End Matter and in the Supplemental Material \cite{SupplementalMaterial}. Throughout, we fix $f/f_c=0.1$, $f_c\simeq0.75$, deep in the creep regime, thereby varying the temperature $T$ without changing the force-controlled activation scale $\ell_{\rm opt}(f)$. A configuration is metastable if every local move $u(i)\to u(i)\pm 1$ increases its energy. At $T\to0^+$, the optimal event, \ie the smallest compact energy-lowering rearrangement, is identified through Dijkstra-based minimization~\cite{Ferrero2017} and followed by deterministic relaxation.

At finite temperature, rather than selecting only the optimal rearrangement, we consider multiple competing forward rearrangements that lower the energy. For each segment starting at $i$ with length $\ell$, we compute the energy-minimizing path between $u(i)$ and $u(i+\ell)$ using Dijkstra's algorithm. If the resulting path differs from the current configuration, it defines a candidate rearrangement. Each such candidate is assigned an effective energy barrier $U(\ell)=\ell^{1/3}$, consistent with equilibrium droplet scaling in one dimension~\cite{Drossel1995,Kolton2005}, leading to an Arrhenius rate
\begin{equation}
r(\ell)=\exp\!\left(-\frac{U(\ell)}{T}\right).
\end{equation}

We restrict rearrangements to lengths $\ell\le\ell_{\max}$ and retain a reduced candidate set, as detailed in the End Matter. A kinetic Monte Carlo (KMC) scheme selects a thermal excitation among these candidates with probability proportional to its rate; deterministic VMC relaxation then leads to a new metastable configuration. 
At the small force considered here, this relaxation produces only a small adjustment to the excitation, as quantified in the Supplemental Material. A thermal avalanche is neither a single excitation nor its subsequent deterministic relaxation, but a correlated spatiotemporal sequence of many excitation-relaxation events~\cite{Ferrero2021,Ferrero2017}.

\begin{figure*}[!t]
\centering

\begin{minipage}[t]{0.49\textwidth}
    \centering
    \includegraphics[width=\textwidth]{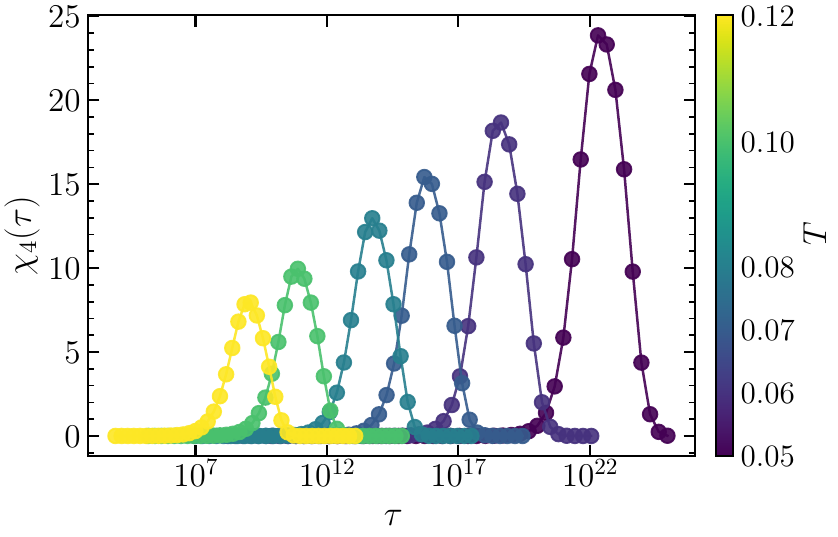}
\end{minipage}
\hfill
\begin{minipage}[t]{0.49\textwidth}
    \centering
    \includegraphics[width=\textwidth]{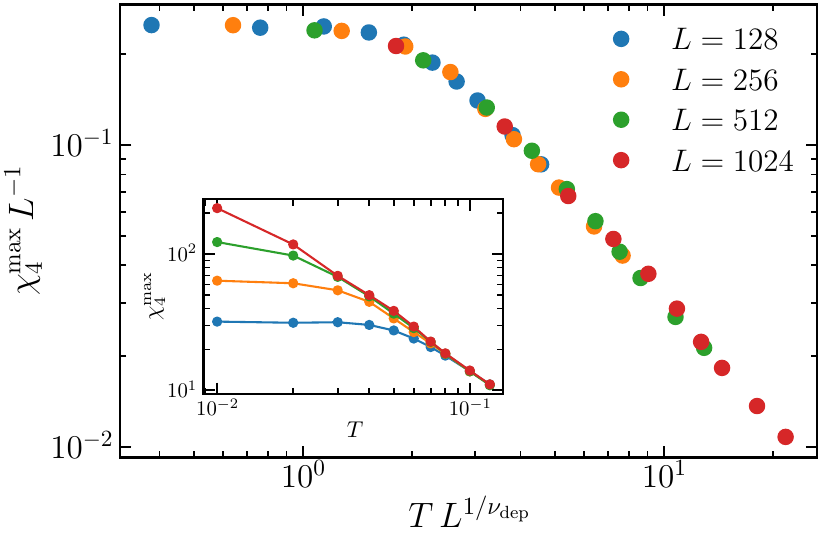}
\end{minipage}

\caption{
Dynamical correlations and avalanche scaling at finite temperature. 
Left: four-point dynamical susceptibility $\chi_4(\tau)$ at fixed drive $f/f_c = 0.1$ and system size $L=512$ for different temperatures. As temperature decreases, the peak of $\chi_4(\tau)$ grows, indicating increasing dynamical correlations associated with thermal avalanches. 
Right: finite-size scaling of the maximum $\chi_4^{\mathrm{max}}$. The main panel shows the collapse of $\chi_4^{\mathrm{max}} L^{-1}$ as a function of $T L^{1/\nu_{\mathrm{dep}}}$, while the inset displays the corresponding raw data. The scaling is consistent with $\chi_4^{\mathrm{max}} \sim \ell_{\mathrm{av}}(T) \sim T^{-\nu_{\mathrm{dep}}}$.
}
\label{fig:fig3}
\end{figure*}

\begin{table}[t]
\centering
\begin{tabular}{|c|c | c| }
\hline
Regime & Exponents & Scales \\
\hline
Equilibrium  
& $\zeta_{\mathrm{eq}}=2/3$, $\nu_{\mathrm{eq}}=3/4$ & small
 \\

Depinning 
& $\zeta_{\mathrm{dep}}=5/4$, $\nu_{\mathrm{dep}}=4/3$ &intermediate
\\

Thermal 
& $\zeta_{\mathrm{th}}=1/2$ & large
\\

\hline
\end{tabular}
\caption{Roughness regimes and associated critical exponents for a driven elastic line. The relation $\nu = 1/(2-\zeta)$ holds both at equilibrium and at depinning~\cite{Rosso2003, Ferrero2013, Shapira2023}.}
\label{tab:exponents}
\end{table}

\emph{Results.}
We characterize the interface geometry through its structure factor, a scale-dependent measure of roughness. The Fourier transform is defined as
\begin{equation*}
u(q)=\sum_{j=0}^{L-1}e^{-\mathrm{i}qj}u(j),
\qquad q=\frac{2\pi n}{L},
\end{equation*}
and the structure factor as
\begin{equation}
S(q)=\overline{|u(q)|^2},
\end{equation}
where the overbar denotes an average over steady-state metastable configurations. For a self-affine interface, $S(q)\sim q^{-(1+2\zeta)}$, where $\zeta$ is the roughness exponent on scale $q^{-1}$.

In the limit $T\to0^+$, the interface crosses over from equilibrium roughness, $\zeta_{\mathrm{eq}}=2/3$, at short length scales to depinning roughness, $\zeta_{\mathrm{dep}}=5/4$, at larger scales~\cite{Kolton2009,Ferrero2017}. The crossover scale, $\ell_{\mathrm{opt}}(f)\sim f^{-\nu_{\mathrm{eq}}}$, with $\nu_{\mathrm{eq}}=1/(2-\zeta_{\mathrm{eq}})=3/4$, remains essentially unchanged at finite temperature, as shown in Fig.~\ref{fig:fig1}. The equilibrium roughness observed below $\ell_{\mathrm{opt}}$ also provides an \emph{a posteriori} validation of the equilibrium droplet scaling assumed for the energy barriers. 

At larger scales, finite temperature induces a second crossover at the avalanche scale $\ell_{\mathrm{av}}(f,T)$, distinct from $\ell_{\mathrm{opt}}(f)$: depinning roughness persists for $\ell_{\mathrm{opt}}<q^{-1}<\ell_{\mathrm{av}}$, while at still larger scales the interface approaches thermal roughness with $\zeta_{\mathrm{th}}=1/2$ (Table~\ref{tab:exponents}).

At fixed force, rescaling wavevectors by $T^{-\nu_{\mathrm{dep}}}$ collapses the data, yielding
\begin{equation}
\ell_{\mathrm{av}}(f,T)\sim T^{-\nu_{\mathrm{dep}}},
\qquad
\nu_{\mathrm{dep}}=\frac{1}{2-\zeta_{\mathrm{dep}}}.
\end{equation}
The thermal regime also collapses because its roughness amplitude is set by disorder-induced effective noise, whereas temperature only sets the crossover scale $\ell_{\mathrm{av}}$. Thermal roughness should therefore emerge asymptotically at small $q$ whenever $L\gg\ell_{\mathrm{av}}$.

Several approaches predict different scalings for the avalanche cutoff. Functional renormalization group arguments give $\ell_{\mathrm{av}}\sim T^{-\nu_{\mathrm{dep}}/\beta_{\mathrm{dep}}}$~\cite{Chauve2000,Chauve2001}. The scaling argument of Ref.~\cite{Tahaei2023} predicts $\ell_{\mathrm{av}}\sim T^{-1/(d\delta)}$, where $\delta$ is defined through the scaling of the typical gap between the two lowest activation barriers in a region of size $\ell$, $\Delta E_{\mathrm{gap}}\sim \ell^{-d\delta}$. For depinning, the monotonic elastic kernel produces no pseudogap in the distribution of local stabilities, so that $\delta=1$; in $d=1$, this yields $\ell_{\mathrm{av}}\sim T^{-1}$.
By contrast, Ref.~\cite{deGeus2024} predicts $\ell_{\mathrm{av}}\sim T^{-\nu_{\mathrm{dep}}}$, directly controlled by the depinning correlation-length exponent. Our results support this scenario. 

We now discuss the system's dynamics. We use two standard probes of dynamical heterogeneities in glassy systems~\cite{Berthier2011}: the persistence and the four-point dynamical susceptibility $\chi_4$. The persistence is
\begin{equation}
\pi(\tau)=\frac{1}{L}\sum_{i=0}^{L-1}p_i(\tau),
\end{equation}
where $p_i(\tau)=1$ if site $i$ remains immobile during $\tau$, and $p_i(\tau)=0$ otherwise. As $T$ decreases, $\langle\pi(\tau)\rangle_{\mathrm{time}}$ decays more slowly. Defining $\tau_\alpha$ by $\langle\pi(\tau_\alpha)\rangle_{\mathrm{time}}=0.5$, we find Arrhenius behavior (Fig.~\ref{fig:fig2}), showing that the dynamical timescale is controlled by large, temperature-independent barriers associated with rearrangements of size $\ell_{\mathrm{opt}}$, the activation bottleneck.

We quantify spatial dynamical correlations through the four-point susceptibility
\begin{equation}
\chi_4(\tau)=L\left(\langle\pi^2(\tau)\rangle_{\mathrm{time}}
-\langle\pi(\tau)\rangle_{\mathrm{time}}^2\right).
\end{equation}
As $T$ decreases, $\chi_4(\tau)$ develops a growing peak (Fig.~\ref{fig:fig3}, left), signaling larger dynamically correlated regions; its maximum, $\chi_4^{\mathrm{max}}$, estimates their size.
One-dimensional depinning avalanches are spatially compact: every site in the affected interval is activated at least once, so this size scales as $\ell_{\mathrm{av}}$.
We find (Fig.~\ref{fig:fig3}, right)
\begin{equation}
\chi_4^{\mathrm{max}}(T)\sim T^{-\nu_{\mathrm{dep}}},
\end{equation}
in agreement with the fixed-$f$ scaling of $\ell_{\mathrm{av}}$ obtained from $S(q)$ (Fig.~\ref{fig:fig1}). This independently confirms that $\ell_{\mathrm{av}}$ sets the spatial extent of collective motion.

\emph{Conclusions.}
We have identified two distinct length scales in finite-temperature creep of a driven elastic line: a temperature-independent activation scale $\ell_{\mathrm{opt}}(f)$, associated with the bottleneck controlling the dynamical timescale, and a thermal-avalanche scale $\ell_{\mathrm{av}}(f,T)\sim T^{-\nu_{\mathrm{dep}}}$ at fixed $f$, which controls the spatial organization of the dynamics.

Thermal avalanches have been reported in systems ranging from magnetic domain walls to crumpling experiments~\cite{Durin2024, Lahini2023}, as well as in elastoplastic models of amorphous solids~\cite{Tahaei2023, Ozawa2023, Rodriguezlopez2026}. The analogy with amorphous solids remains qualitative, since yielding and depinning involve different interaction kernels and universality classes.

An intriguing connection concerns dynamical heterogeneities in glassy systems~\cite{Ciamarra2026}, where the spatial extent of dynamical correlations grows upon cooling~\cite{Berthier2005, Karmakar2009}. The physical origin of this growth remains debated, with proposed explanations ranging from random first-order transition theory and kinetically constrained models to approaches based on elastic interactions and marginal stability~\cite{Biroli2009, Garrahan2002, Wyart2017, Lemaitre2014}. Our results suggest testing whether this growth can be described by a thermal-avalanche scale. However, our fixed-drive study does not address the strictly undriven limit: while $\ell_{\mathrm{opt}}(f)$ diverges as $f\to0^+$ for elastic interfaces, whether an analogous scale exists in amorphous and glassy systems remains open. Recent mean-field studies of activated jumps in energy landscapes provide a promising starting point for exploring this connection~\cite{pacco2025triplets}.

\acknowledgements
\emph{Acknowledgements.} 
We warmly thank L. Foini for insightful support. We also thank T. de Geus, M. Ghizzoni, T. Giamarchi, V. M. Schimmenti, J. Weiss, and M. Wyart for useful discussions. This work was supported by the French National Research Agency (ANR), project DISCREEP (ANR-23-CE30-0031-04), and by the CNRS IRP ``Statistical Physics of Materials''.

\bibliographystyle{apsrev4-2}
\bibliography{library,creep_exp}

\begin{figure*}[t]
\centering
\includegraphics[width=\textwidth]{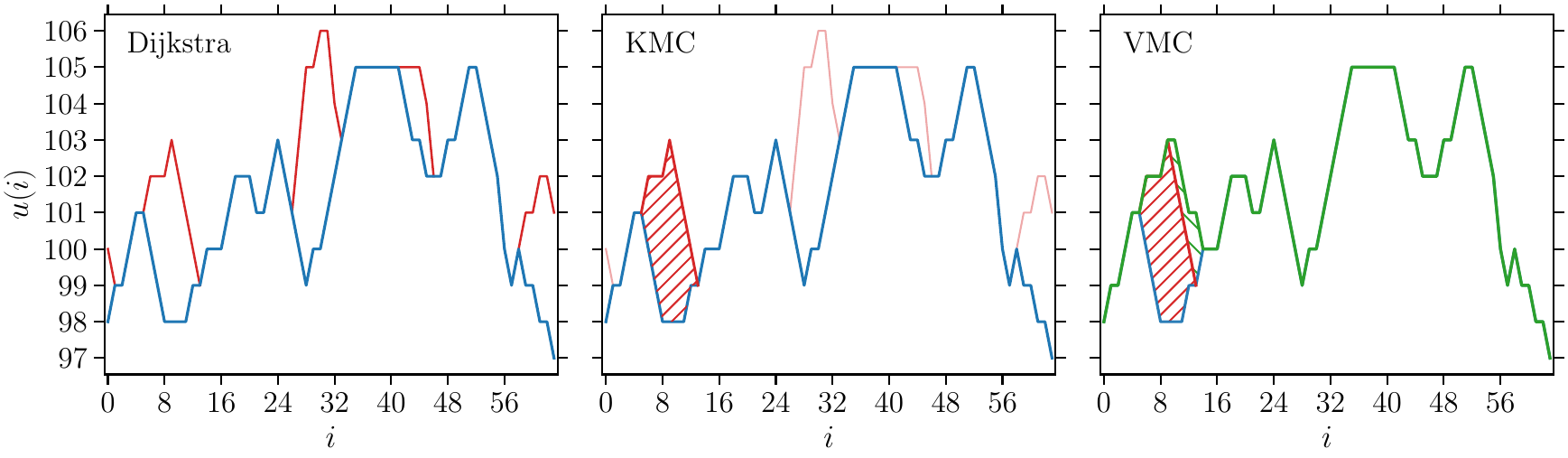}
\caption{
Schematic illustration of the finite-temperature dynamics. 
Left: identification of energy-lowering rearrangements (red) from a metastable configuration (blue) using a Dijkstra-based search. 
Middle: stochastic selection of an activated event via kinetic Monte Carlo (KMC), with probabilities set by Arrhenius activation rates. 
Right: deterministic relaxation using variant Monte Carlo (VMC), yielding a new metastable configuration (green).
}
\label{fig:reorganizations}
\end{figure*}

\onecolumngrid 

\bigskip \bigskip
\centerline {\bf End Matter}
\bigskip 

\twocolumngrid

\appendix

\section{Algorithm}

We describe the numerical procedure used to simulate the finite-temperature dynamics. The method extends the algorithm introduced in Ref.~\cite{Ferrero2017} and is schematically illustrated in Fig.~\ref{fig:reorganizations}. Starting from a metastable configuration, each dynamical update consists of three steps: (i) identification of candidate rearrangements, (ii) stochastic selection via kinetic Monte Carlo, and (iii) deterministic relaxation toward a new metastable configuration.

We consider a directed polymer with integer displacements $u(i)$. A configuration is metastable if any local move $u(i)\to u(i)\pm 1$ increases its energy.

\paragraph{(i) Identification of candidate rearrangements}

Starting from a metastable configuration, we identify forward rearrangements that lower the energy. For each segment of starting point $i$ and length $\ell$, we compute the optimal path between $i$ and $i+\ell$ using a Dijkstra-based algorithm. A rearrangement is retained whenever this path differs from the current configuration and leads to a lower-energy state.

We have tested two different prescriptions to construct the set of candidate rearrangements. In both cases, we restrict to rearrangements up to a cutoff $\ell_{\max}=50$. Rearrangements at larger scales have vanishingly small activation probabilities and, therefore, do not contribute to the dynamics. 

In the first prescription, all energy-lowering rearrangements up to this cutoff are included. In the second, rearrangements are selected hierarchically, starting from the smallest ones and progressively adding larger ones under a non-overlapping constraint, resulting in a reduced set of candidates.

We have verified that these two implementations yield results that are statistically indistinguishable within numerical accuracy. In the main text, we present results obtained using the second prescription.

\paragraph{(ii) Activation rates and event selection}

Each rearrangement of size $\ell$ is assigned an effective energy barrier
$U(\ell) = \ell^{1/3}$, consistent with equilibrium droplet scaling in one dimension \cite{Drossel1995, Kolton2005}. This defines an Arrhenius rate
\begin{equation}
r(\ell) = \exp\!\left(-\frac{U(\ell)}{T}\right).
\end{equation}

An activated event is selected among the candidates with probability proportional to its rate, using a kinetic Monte Carlo (KMC) scheme. 
The KMC procedure also advances the physical time according to the total escape rate from the current metastable configuration,
{
\begin{equation}
\Delta t=-\frac{\ln \xi}{\sum_\alpha r_\alpha},
\end{equation}
}
where $\xi$ is uniformly distributed in $(0,1)$ and the sum runs over all candidate excitations.

\paragraph{(iii) Deterministic relaxation}

After the activated move, the system need not be in a metastable state and may relax deterministically. We use the variant Monte Carlo (VMC) dynamics of Ref.~\cite{Rosso2001}, which allows for the simultaneous motion of $k+1$ adjacent sites by one lattice spacing if no move of $k$ sites is energetically favorable. This choice avoids pathologies associated with single-site dynamics and ensures efficient relaxation toward metastable configurations.

The thermally activated rearrangement selected in step (ii) constitutes an elementary thermal excitation, while the subsequent VMC relaxation completes the transition to the next metastable configuration. A thermal avalanche does not correspond to either this deterministic relaxation or a single excitation, but consists of a correlated spatiotemporal sequence of many elementary thermal excitations~\cite{Ferrero2021,Ferrero2017}.

We neglect the duration of this deterministic relaxation compared with the thermally activated waiting time. This separation is justified in the low-temperature creep regime, where activation times are exponentially large while the VMC relaxation occurs on a microscopic, non-activated timescale. At the small driving force considered here, the dense sequence of metastable configurations also makes the deterministic relaxation spatially small. A quantitative test is reported in the Supplemental Material, where we also test the influence of the metric constraint by comparing simulations with $K=1,\dots,5$.

\paragraph{Steady state}

We focus on steady-state properties. The initial transient is discarded by monitoring the interface width and restricting the analysis to times when it has saturated, ensuring that no systematic growth remains~\cite{barabasi1995fractal}.

\paragraph{Limit $T\to0^+$}

In the limit $T\to0^+$, the dynamics reduces to the selection of the rearrangement with the smallest barrier, which corresponds to the smallest compact rearrangement that lowers the energy. This recovers the algorithm of Ref.~\cite{Ferrero2017}. This limit has served as a benchmark for the dynamics, as discussed in the following section.

\begin{figure}[ht!]
\centering
\includegraphics[width=\columnwidth]{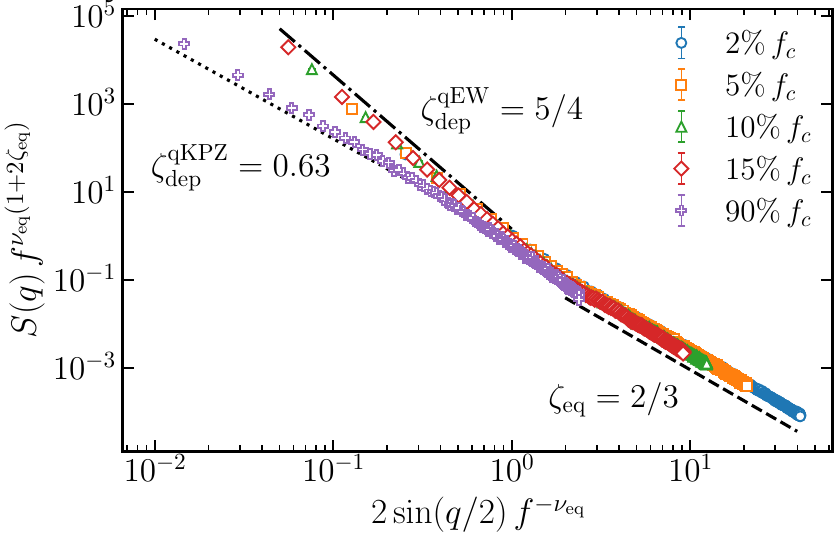}
\caption{
Structure factor $S(q)$ at different driving forces in the limit $T \to 0^+$. 
For each force, the structure factor is averaged over $10^3$ configurations.
A crossover from the equilibrium roughness exponent $\zeta_{\mathrm{eq}}=2/3$ to the quenched Edwards--Wilkinson depinning exponent $\zeta_{\mathrm{dep}}^{\mathrm{qEW}}=5/4$ is observed at small forces. Only close to the critical force ($f/f_c = 0.9$) does a further crossover toward the qKPZ exponent $\zeta_{\mathrm{dep}}^{\mathrm{qKPZ}}\simeq 0.63$ appear.
}
\label{fig:collapsezeroplus}
\end{figure}

\section{Structure factor in the $T \to 0^+$ limit}
\label{sec:qKPZ}

The depinning transition of the model with a hard metric constraint ($K=1$) belongs to the qKPZ universality class. However, as previously observed in Ref.~\cite{Kolton2009}, the nonlinearity induced by the constraint produces a very large crossover scale.

Figure~\ref{fig:collapsezeroplus} shows the structure factor obtained at different driving forces in the $T\to 0^{+}$ limit. At small driving forces, a clear crossover is observed between the equilibrium roughness exponent $\zeta_{\rm eq}=2/3$ and the quenched Edwards-Wilkinson depinning exponent $\zeta_{\rm dep}=5/4$.

Only for a driving force close to the critical value,
$f/f_c = 0.9$, do the data exhibit a further crossover
toward the expected qKPZ exponent
$\zeta_{\mathrm{dep}}^{\mathrm{qKPZ}}\simeq 0.63$.
This qEW-qKPZ crossover has been previously studied in the limit of vanishing temperature~\cite{Ferrero2021}. It is controlled by an anharmonic crossover length which, for our parameters, is larger than the thermal avalanche scale $\ell_{\rm av}(f, T)$, thereby hiding the qKPZ regime.
This justifies our use of the quenched Edwards-Wilkinson depinning exponents in our finite-temperature simulations.
The proximity between the qKPZ and thermal exponents, $\zeta_{\rm dep}^{\rm qKPZ} \simeq 0.63$ and $\zeta_{\rm th} = 1/2$, is merely numerical.

\end{document}